\documentstyle[epsfig]{article}

\input{epsf}

\begin{document}

\title{
{\noindent\small UNITU--THEP--1/1999 \hfill FAU--TP3--99/1 
\hfill hep-ph/9901322 }\\
The Infrared Behavior of Gluon, Ghost, and Quark Propagators 
in Landau Gauge QCD}

\author{Reinhard Alkofer, Steven Ahlig\\[5pt]
         Universit\"at T\"ubingen, Institut f\"ur Theoretische Physik,\\
         Auf der Morgenstelle 14,
         72076 T\"ubingen, Germany\\
	 E-mail: Reinhard.Alkofer@uni-tuebingen.de\\
	 \phantom{E-mail:} ahlig@pthp5.tphys.physik.uni-tuebingen.de\\[10pt]
         and Lorenz von Smekal\\[5pt]
         Universit\"at Erlangen--N\"urnberg,\\ 
	 Institut f\"ur Theoretische Physik III,\\
	 Staudtstr.~7, 
         91058 Erlangen, Germany\\
	 E-mail: smekal@theorie3.physik.uni-erlangen.de}

\maketitle	 
	 

\begin{abstract}
A truncation scheme for the Dyson--Schwinger equations of QCD in Landau gauge
is presented which implements the Slavnov--Taylor identities for the 3--point
vertex functions. Neglecting contributions from 4--point correlations such as
the 4--gluon vertex function and irreducible scattering kernels, a closed
system of equations for the propagators is obtained. For the pure gauge
theory without quarks this system of equations for the propagators of gluons
and ghosts is solved in an approximation which allows for an analytic
discussion of its solutions in the infrared: The gluon propagator is shown
to vanish for small spacelike momenta whereas the ghost propagator is found
to be infrared enhanced. The running coupling of the non--perturbative
subtraction scheme approaches an infrared stable fixed point at a
critical value of the coupling, $\alpha_c \simeq 9.5$. The gluon propagator
is shown to have no Lehmann representation. The gluon and ghost propagators
obtained here compare favorably with recent lattice calculations. 
Results for the quark propagator in the quenched approximation are presented.

\end{abstract}

\section{Introduction}

Despite the remarkable success of perturbative QCD the description of hadronic
states and processes based on the dynamics of confined  quarks and gluons remains
the outstanding challenge of strong interaction physics. Especially, one has to
explain why only hadrons are produced from processes involving hadronic
initial states, and that the only thresholds in hadronic amplitudes are due to
the productions of other hadronic states. To this end one would like to understand
how singularities appear in the Green's functions of composite hadron fields
where, on the other hand, they have to disappear in colored correlations
functions. 

To study these aspects of QCD amplitudes non--perturbative methods are
required, and, since infrared divergences are anticipated, a formulation in the
continuum is desirable. Both of these are provided by studies of truncated
systems of Dyson--Schwinger equations (DSEs), the equations of motion of QCD
Green's functions. Typically, for their truncation, additional sources of
information like the Slavnov--Taylor identities, entailed by gauge invariance,
are used to express vertex functions in terms of the elementary two--point
functions, i.e., the quark, ghost and gluon propagators. Those propagators can
then be obtained as selfconsistent solutions to non--linear integral equations
representing a closed set of truncated DSEs. Some systematic control over the
truncating assumptions can be obtained by successively including higher
$n$--point functions in selfconsistent calculations, and by assessing their
influence on lower $n$--point functions in this way. Until recently all
solutions to truncated DSEs of QCD in Landau gauge, even in absence of quarks,
relyed on neglecting ghost contributions completely 
\cite{Man79,Atk81,Bro89,Hau96}. 

In addition to providing a better understanding of con\-fine\-ment based on
studies of the behavior of QCD Green's functions in the infrared, DSEs have
proven successful in developing a hadron phenomenology which
interpolates smoothly between the infrared non--perturbative and the
ultraviolet perturbative regime \cite{Dub98}, for recent reviews see,
{\it e.g.}, \cite{Tan97,Rob94}. In particular, a dynamical description of
spontaneous breaking of chiral symmetry from studies of the DSE for the
quark propagator is well established in a variety of models for the
gluonic interactions of quarks~\cite{Miranski}. For a sufficiently large
low--energy quark--quark interaction quark masses are generated dynamically
in the quark DSE in some analogy to the gap equation in superconductivity.
This in turn leads naturally to the Goldstone nature of the pion and explains
the smallness of its mass as compared to all other hadrons.
In this framework a description of the different types of mesons is obtained
from Bethe--Salpeter equations for quark--antiquark bound states~\cite{Mun92}.
Recent progress towards a solution of a fully relativistic three--body
equation extends this consistent framework to baryonic bound states,
see {\it e.g.} \cite{Oet98} and references therein.

Here a simultaneous solution of a truncated set of DSEs for the propagators of
gluons and ghosts in Landau gauge is presented \cite{Sme97,Sme98}. An extension
of this selfconsistent framework to include quarks is subject to on--going
research \cite{Ahl99}. Preliminary results for the quark propagator in the
quenched approximation have been obtained and will be shown. The behavior of
the solutions in the infrared, implying the existence of a fixed point at a
critical coupling $\alpha_c \approx 9.5$, is obtained analytically. The gluon
propagator is shown to vanish for small spacelike momenta in the present
truncation scheme. This behavior, though in contradiction with previous DSE
studies~\cite{Man79,Atk81,Bro89,Hau96}, can be understood from the observation
that, in our present calculation, the previously neglected ghost propagator
assumes an infrared enhancement similar to what was then obtained for the
gluon. In the meantime such a qualitative behavior of gluon and ghost
propagators is supported by investigations of the coupled gluon ghost DSEs
using bare vertices~\cite{Atk97,Atk98}. As expected, however, the details of
the results depend on the approximations employed.

\section{The set of truncated gluon and ghost DSEs}

Besides all elementary 2--point functions, i.e., the quark, ghost and gluon
propagators, the DSE for the gluon propagator also involves the 3-- and
4--point vertex functions which obey their own DSEs. These equations involve
successively higher n--point functions. The gluon equation is truncated by
 neglecting all terms with 4--gluon vertices. These are
the momentum independent tadpole term, an irrelevant constant which vanishes
perturbatively in Landau gauge, and explicit 2--loop contributions to the gluon
DSE. For all details regarding this truncation scheme we refer the reader
to \cite{Sme98}. 

The ghost and gluon propagators are
parameterized by their respective renormalization functions $G$ and $Z$,
\begin{equation}
  D_G(k) = -\frac{G(k^2)}{k^2} \; , \quad
  D_{\mu\nu}(k) = \bigg( \delta_{\mu\nu} - \frac{k_\mu k_\nu}{k^2}\bigg)
                  \frac{Z(k^2)}{k^2} \; . 
  \label{rfD}
\end{equation}
In order to arrive at a closed set of equations for the functions $G$ and $Z$,
we use a form for the ghost--gluon vertex which is based on a construction from
its Slavnov--Taylor identity (STI) which can be derived from the usual
Becchi--Rouet--Stora invariance neglecting irreducible 4--ghost correlations in
agreement with the present level of truncation~\cite{Sme98}. This together with
the crossing symmetry of the ghost--gluon vertex fully determines its form at 
the present level of truncation:
\begin{equation}
  G_\mu(p,q) =
    i q_\mu \, \frac{G(k^2)}{G(q^2)}
    + i p_\mu \, \biggl( \frac{G(k^2)}{G(p^2)} - 1 \biggr) .
  \label{fvs}
\end{equation}
With this result, we can construct the 3--gluon vertex according to procedures
developed and used previously~\cite{Bar80}, for details see \cite{Sme98}.

We have solved the coupled system of integral equations of the present 
truncation scheme numerically using an angle approximation. The infrared 
behavior of the propagators can, however, be deduced analytically.
To this end we make the Ansatz
that for $x := k^2 \to 0$ the product $Z(x)G(x) \to c x^\kappa$ with $\kappa
\not= 0$ and some constant $c$. The special case $\kappa = 0$ leads to a
logarithmic singularity for $x \to 0$ which precludes
the possibility of a selfconsistent solution. In order to obtain a positive
definite function $G(x)$ for positive $x$ from an equally positive $Z(x)$, as
$x\to 0$, we obtain the further restriction $0 < \kappa <
2$. The ghost DSE then yields, 
\begin{eqnarray}  
  G(x) \to   \left( g^2\gamma_0^G \left(\frac{1}{\kappa} - \frac{1}{2}
  \right) \right)^{-1}  c^{-1} x^{-\kappa} \quad \Rightarrow 
  Z(x)  \to   \left( g^2\gamma_0^G \left(\frac{1}{\kappa} - \frac{1}{2}
  \right) \right) \, c^{2} x^{2\kappa}   \; ,
  \nonumber
\end{eqnarray}
where $\gamma_0^G = 9/(64\pi^2)$ is the leading order perturbative coefficient
of the anomalous dimension of the ghost field. Using these relations
 in the gluon DSE, we find that the 3--gluon loop
contributes terms $\sim  x^\kappa$ to the gluon equation for $x \to 0$ while the
dominant (infrared singular) contribution arises from the ghost--loop, 
\begin{displaymath}
  Z(x) \to g^2\gamma_0^G \, \frac{9}{4} \left(\frac{1}{\kappa} -
  \frac{1}{2} \right)^2 \! \left( \frac{3}{2}\, \frac{1}{2-\kappa} -
  \frac{1}{3} + \frac{1}{4\kappa} \right)^{-1}\!\! c^2 x^{2\kappa} . 
\end{displaymath}
Requiring a unique behavior for $Z(x)$ 
we obtain a quadratic equation for $\kappa$
with a unique solution for the exponent in $0 < \kappa < 2$:
\begin{equation}
  \kappa = \frac{61-\sqrt{1897}}{19} \simeq 0.92 \; .
\end{equation}
The leading behavior of the gluon and ghost renormalization
functions and thus of their propagators is entirely due to ghost contributions.
The details of the approximations to the 3--gluon loop have no influence on the
above considerations. Compared to the Mandelstam approximation, in which the 
3--gluon loop alone determines the infrared behavior of the gluon propagator
and the running coupling in Landau gauge~\cite{Man79,Atk81,Bro89,Hau96}, this
shows the importance of ghosts. The result presented here implies an infrared
stable fixed point in the non--perturbative running coupling of our subtraction
scheme, defined by 
\begin{equation}
  \alpha_S(s) = \frac{g^2}{4\pi} Z(s) G^2(s)
    \quad \to \quad
    \frac{16\pi}{9} \left(\frac{1}{\kappa} - \frac{1}{2}\right)^{-1}
    \approx 9.5 \; , 
\end{equation} 
for $s\to 0$. This is qualitatively different from the infrared singular 
coupling of the Mandelstam approximation~\cite{Hau96}.

\section{Comparison to lattice results}

It is interesting to compare our solutions to recent lattice results available
for the gluon propagator~\cite{Der98} and for the ghost propagator~\cite{Sum96}
using lattice versions to implement the Landau gauge condition. We would like
to refer the reader to ref.\ \cite{Hau98} where this has been done in some
detail. It is very encouraging to
observe that our solution fits the lattice data at low momenta rather well,
especially for the ghost propagator.
We therefore conclude that present lattice calculations
confirm the existence of an infrared enhanced ghost propagator of the form $D_G
\sim 1/(k^2)^{1+\kappa}$ with $0 < \kappa < 1$. This is an interesting result
for yet another reason: In the calculation of \cite{Sum96}
the Landau gauge condition was
supplemented by an algorithm to select  gauge field configurations from the
fundamental modular region which is to avoid Gribov copies. Thus, our results
suggest that the existence of such copies of gauge configurations might have
little effect on the solutions to Landau gauge DSEs. 

\begin{figure}[t]
\vskip .6cm
\hskip -.6cm
\parbox{4cm}{
  \epsfig{file=Bou98.eps,width=5.8cm}}
\hskip 1.7cm
\parbox{4cm}{
\vskip -1cm
  \epsfig{file=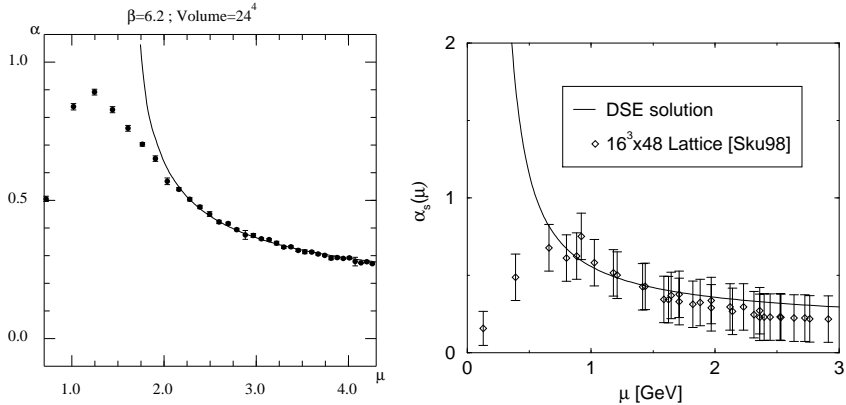,width=5.8cm}}
\vskip -.4cm  
  \caption{Lattice results of the running coupling from the 3--gluon vertex
  (left, together with a 3--loop fit, Fig.~1 of Ref.~\cite{Bou98}), and from
  the quark--gluon vertex for $\beta = 6.0$ on a $16^3 \times 48$ lattice
  (right, c.f.\ Fig.~2 of Ref.~\cite{Sku98}).}
  \label{lat_alpha}
\end{figure}

Here we want to add a remark concerning the comparison of the running coupling
obtained in our calculation to lattice results.
Recent lattice calculations of the running coupling are reported in
Refs.~\cite {All97,Bou98} based on the 3--gluon vertex, and Ref.~\cite{Sku98}
on the quark--gluon vertex. The non--perturbative definitions of these
couplings are related but manifestly different from the one adopted here.
One of the most recent results from the 3--gluon vertex is shown in the left 
graph of Fig.~\ref{lat_alpha}  and compared to the three--loop expression
which is for the momenta displayed almost identical to our expression (4) for
the running coupling.
This lattice result is obtained from an asymmetric momentum
subtraction scheme. This corresponds to a definition of the
running coupling $\bar g^2_{3GVas}$ which can explicitly be related to the
present one ($\bar g^2(t,g)$ with $t = \ln \mu'/\mu$ and $g := g(\mu)$),
\begin{equation}
\bar g^2(t,g^2)_{3GVas} \, = \, \bar g^2(t,g^2) \, \lim_{s \to 0} \,
\frac{G^2(s)}{G^2({\mu'}^2)} \, \left( 1 \, - \, \frac{\beta(\bar
g(t,g))}{\bar g(t,g)}
\, \right)^2 \; . \label{comp_rc}
\end{equation}
An inessential difference in these two definitions of the running coupling
is the last factor in brackets in eq.~(\ref{comp_rc}) which can be easily
accounted for in comparing the different schemes. 
However, the crucial difference is the ratio
of ghost renormalization functions $G(s\to 0)/G({\mu'}^2)$. These 
considerations show that
the asymmetric scheme can be extremely dangerous if infrared divergences
occur in vertex functions as our calculation indicates. 
Clearly, from the infrared enhanced ghost renormalization function this scale
dependence could account for the infrared suppressed couplings which seem to
be found in the asymmetric schemes.

Similarly, the results from the quenched calculation of the quark--gluon
vertex of Ref.~\cite{Sku98} which are compared in the right graph of 
Fig.~\ref{lat_alpha} to our solution
are obtained from an analogous asymmetric scheme. 
It is thus expected to have the same problems in taking
the possible infrared divergences of the vertices into account which arise
in both, the 3--gluon and the quark--gluon vertex, as a result of the
infrared enhancement of the ghost propagator. 

Furthermore,  definitions of the
coupling which lead to extremas at finite values of the scale correspond to
double valued $\beta$--functions with artificial zeros. If the maxima in the
couplings of the asymmetric schemes at finite scales 
are no lattice artifacts, these results seem to
imply that the asymmetric schemes are less suited for a non--perturbative
extension of the renormalization group to all scales. Indeed,
the results for the running coupling from the 3--gluon vertex obtained for
the symmetric momentum subtraction scheme in Ref.~\cite{Bou98} differ from
those of the asymmetric scheme, in particular, in the infrared. 
These results would be better to compare to the DSE
solution, however, they unfortunately seem to be much noisier thus far (see
Ref.~\cite{Bou98}).

The ultimate lattice calculation to compare to the present DSE coupling would
be obtained from a pure QCD calculation of the ghost--gluon vertex in Landau
gauge with a symmetric momentum subtraction scheme. This is unfortunately not
available yet.

\section{Quark Propagator}

We have solved the quark DSE in quenched approximation \cite{Ahl99}.
In a first step we have specified the quark--gluon vertex from the corresponding
Slavnov--Taylor identity. It contains explicitely a ghost renormalization
function,
\begin{equation}
  \Gamma^\mu (p,q) = G(k^2) \Gamma^\mu _{\rm CP} (p,q)
\end{equation}
where $\Gamma^\mu _{\rm CP}$ is the Curtis--Pennington vertex (for its
definition see {\it e.g.} \cite{Rob94}). It is obvious that this leads to an
effective coupling very different from the one in Abelian approximation,
especially in the infrared: This effective coupling vanishes in the infrared
and is similar to the lattice result of  Ref.~\cite{Sku98} shown in the right 
graph of  Fig.~\ref{lat_alpha} the main difference being that the maximum occurs
at lower scale, $\mu \approx 220$MeV.
This leads to a kernel in the quark DSE which is only very slightly infrared
divergent. This allows, {\it e.g.}, to use the Landshoff--Nachtmann model
for the pomeron in our approach. With our solution we obtain as Pomeron
intercept 2.7/GeV as compared to the value 2/GeV deduced from phenomenology,
see {\it e.g.} \cite{Cud89}. 

We have found dynamical chiral symmetry breaking in the quenched approximation.
Using a current mass, $m(1{\rm GeV})=6$MeV we obtain a constituent mass of
approximately 170 MeV. In the Pagels--Stokar approximation the calculated value 
for the pion decay constant is 50 MeV. These numbers are quite encouraging,
especially for proceeding with the self--consistent inclusion of the quark DSE
into the gluon--ghost system.

Considering the quark loop in the gluon DSE one realizes that the quark loop
will produce an infrared divergence which is, however, subleading as compared to
the one generated by the ghost loop. In the latter there appear three ghost
renormalization functions in the numerator and one in the denominator leading
effectively to an infrared divergence of the order $(k^2)^{-2\kappa}$. In the
quark loop term there is only one factor $G$ and thus a divergence of type
$(k^2)^{-\kappa}$. Due to this subleading divergence the infrared analysis
has to be redone completely before one is able to draw conclusions whether or not and
how quark confinement is implemented in our set of truncated DSEs.

\section{Summary}

In summary, we presented a solution to a truncated set of coupled
Dyson--Schwinger equations for gluons and ghosts in Landau gauge. The infrared
behavior of this solution, obtained analytically, represents a strongly
infrared enhanced ghost propagator and an infrared vanishing gluon propagator.


The Euclidean gluon correlation function presented here can be shown to violate
reflection positivity~\cite{Sme98}, which is a necessary and sufficient
condition for the existence of a Lehmann representation. We
interpret this as representing confined gluons. In order to understand how
these correlations can give rise to confinement of quarks, it will be necessary
to redo the infrared analysis including self--consistently the quark propagator.
Nevertheless, we found dynamical chiral symmetry breaking in the quenched
approximation.

The existence of an infrared fixed point for the coupling is in
qualitative disagreement with previous studies of the gluon DSE neglecting
ghost contributions in Landau gauge~\cite{Man79,Atk81,Bro89,Hau96}. 
On the other hand, our results for the propagators, in particular for the
ghost, compare favorably with recent lattice calculations~\cite{Der98,Sum96}.
This shows
that ghosts are important, in particular, at low energy scales relevant to
hadronic observables.

\section*{Acknowledgments}

R.\ A.\ wants to thank the organizers of the conference, and especially
Dubravko Klabucar, for their help and the warm hospitality extended to him.

\noindent
This work was supported by DFG (Al 279/3-1) and the BMBF (06--ER--809).

\end{document}